\documentstyle[12pt]{article}
  \hoffset = - 1 truecm
\def\beq{\begin{equation}}  \def\eeq{\end{equation}}
\def\bea{\begin{eqnarray}}  \def\eea{\end{eqnarray}} \def\nn{\nonumber}
\def\noi{\noindent} \def\beeq{\begin{eqnarray}}
\def\eeeq{\end{eqnarray}}
\def\lsim{\raise0.3ex\hbox{$<$\kern-0.75em\raise-1.1ex\hbox{$\sim$}}}
\def\gsim{\raise0.3ex\hbox{$>$\kern-0.75em\raise-1.1ex\hbox{$\sim$}}}

\newcommand\mysection{\setcounter{equation}{0}\section}
\renewcommand{\theequation}{\thesection.\arabic{equation}}
\newcounter{hran} \renewcommand{\thehran}{\thesection.\arabic{hran}}

\def\bmini{\setcounter{hran}{\value{equation}}
\refstepcounter{hran}\setcounter{equation}{0}
\renewcommand{\theequation}{\thehran\alph{equation}}\begin{eqnarray}}

\def\bminiG#1{\setcounter{hran}{\value{equation}}
\refstepcounter{hran}\setcounter{equation}{-1}
\renewcommand{\theequation}{\thehran\alph{equation}}
\refstepcounter{equation}\label{#1}\begin{eqnarray}}

\def\emini{\end{eqnarray}\relax\setcounter{equation}{\value{hran}}\renewcommand{\theequation}{\thesection.\arabic{equation}}}

\def\ben{\begin{enumerate}}  \def\een{\end{enumerate}}

\def\cite#1{[\ref{#1}]} \def\citd#1#2{[\ref{#1},\ref{#2}]}

\def\citm#1#2{[\ref{#1}--\ref{#2}]}

\begin{document} 
\begin{center} 
\vbox to 1 truecm {} 
{\large \bf Gravitational S-Duality Realized on NUT - Schwarzschild}
\par \vskip 3 truemm 
{\large \bf  and NUT - de Sitter Metrics} \par
\vskip 3 truemm
{\large \bf }
\vskip 1 truecm {\bf U. Ellwanger} 
\vskip 3 truemm

{\it Laboratoire de Physique Th\'eorique\footnote{Unit\'e Mixte de
Recherche CNRS - UMR N$^{\circ}$ 8627},\\ Universit\'e Paris XI,
B\^atiment 210, 91405 Orsay Cedex, France\\ E-mail:
Ulrich.Ellwanger@th.u-psud.fr} 
\end{center} 
\vskip 2 truecm

\begin{abstract} Gravitational $S$-duality is defined by the
contraction of two indices of the Riemann tensor with the $\varepsilon$
tensor. We review its realization in linearized gravity, and study its
generalization to full non-linear gravity by means of explicit
examples: Up to a rescaling of the coordinates, it relates two
Taub-NUT-Schwarzschild metrics by interchanging $m$ with $\ell$,
provided both parameters are non-zero. In the presence of a
cosmological constant gravitational $S$-duality can be implemented at
the expense of the introduction of a three-form field, whose value
turns out to be dual to the cosmological constant. 
\end{abstract} 
\vskip 3 truecm 
\noi LPT Orsay 01-117 \par 
\noi December 2001

\newpage 
\pagestyle{plain} 
\baselineskip=24 pt 
\mysection{Introduction}
\hspace*{\parindent} 
$S$-dualities play an important role in relating different
configurations in string- and $M$-theory. $S$-dualities are best
understood between theories of free abelian $n$-form gauge fields in
arbitrary dimensions. In four dimensional Maxwell theory, for example,
$S$-duality replaces electric charges by magnetic charges and vice
versa. After quantification Dirac's quantization condition then allows
to relate the strong and weak coupling phases of the theory.\par

Given the non-trivial gravitational backgrounds of many configurations
in string- and $M$-theory it becomes important to learn more about
$S$-duality in gravity (even without quantification, i.e. without the
analog of a Dirac quantization condition). An important example is the 
Taub-NUT solution of Einstein's equations \citd{1r}{2r}, which is
characterized by two parameters: a ``Schwarzschild'' mass $m$ and a
``NUT'' parameter $\ell$. The interpretation of $\ell$ as the
gravitational analog of a magnetic charge has a long history \cite{3r}.
This suggests the existence of a duality-like transformation in pure
gravity, and various proposals for such transformations have been made
\citm{4r}{9r}. Not all of them \cite{7r}, however, correspond to
$S$-duality in the sense that the parameters $m$ and $\ell$ of the
Taub-NUT solution get interchanged. \par

At the level of linearized gravity, where the non-abelian structure of
the Lorentz algebra does not yet become apparent, $S$-duality has been
considered recently in \citd{8r}{9r}. In full non-linear gravity the
most natural definition of gravitational $S$-duality is obtained using
the method of differential forms \cite{10r}, where duality acts on the
Riemann tensor with indices in a flat tangent space. This approach has
been used widely for the search for Riemann tensors which are (anti-)
self-dual, since these are automatically solutions of the (Euclidean)
Einstein equations \citd{11r}{10r}. The Euclidean Taub-NUT solution
with $\ell = m$ is a corresponding example \citd{11r}{10r}. \par

The interesting question is in how far $S$-duality exists in full
non-linear gravity in the following sense: Whereas the dual of an
arbitrary Riemann tensor can always be constructed, it is not clear
whether there exists always a dual metric associated to it. (This can
only be proven in linearized gravity, see \citd{8r}{9r} and below.) In
order to find out under which conditions such a dual metric exists it
is helpful to have at one's disposal as many explicit examples as
possible. \par

It is the purpose of the present paper to provide two families of such
examples. First we consider Taub-NUT spaces with $\ell \not= m$. We
find that a dual metric can be constructed explicitly, which allows us
to discuss its properties. We find indeed that, up to a rescaling of
the metric (or the coordinates), the dual metric is again a Taub-NUT
space with $\ell$ and $m$ interchanged. The required rescaling,
however, involves the ratio $\ell/m$ and becomes singular in the limits
$\ell \to 0$ or $m \to 0$. Hence the dual of a ``pure'' Schwarzschild
metric does not exist. This result is very different from $S$-duality
in four dimensional Maxwell theory, which acts without problems on pure
electric or pure magnetic charges. \par

Next we consider Taub-NUT-de Sitter spaces with $\ell \not= m$ and a
Ricci tensor proportional to a cosmological constant $\Lambda$. We find
that a duality transformation can still be defined, at the price of
introducing an additional three-form field $A_{abc} = A_{[abc]}$ with a
field strength $F_{abcd} = F_{[abcd]}$ (here, for convenience, the
latin letters denote indices in flat tangent space). \par

Such three-form fields do not constitute dynamical degrees of freedom
in four dimensions, but their vacuum configuration $F_{abcd} = \Sigma\
\varepsilon_{abcd}$ (with $\Sigma$ = const.) contributes to and
possibly cancels a cosmological constant \cite{12r}. Here, however,
their role is {\it not} to cancel a cosmological constant, but we find
that they are {\it dual} to a cosmological constant in the sense that
the duality transformation interchanges $\Sigma$ and $\Lambda$. This
result may give a new twist to the cosmological constant problem. \par

The rest of the paper is organized as follows: In the next chapter we
review the dual of linearized gravity. Here a dual metric can always be
defined thanks to a generalization of the Poincar\'e lemma (trivial
cohomology) valid for irreducible tensor fields with mixed symmetries
\cite{13r} as in the case of the Riemann tensor. In chapter three we
switch to full non-linear gravity using duality in differential
geometry, and compute and discuss the duals of Taub-NUT metrics. In
chapter four we introduce a generalization of the duality
transformation for non-vanishing Ricci tensor and in the presence of a
three-form field, and generalize the previous results to Taub-NUT-de
Sitter metrics. In chapter five we conclude with an outlook.

\mysection{The Dual of Linearized Gravity} \hspace*{\parindent} In
linearized gravity the deviation of the metric tensor $g_{\mu\nu}$ from
the flat Minkowski metric $\eta_{\mu\nu}$ is considered to be small:

\beq \label{2.1e} g_{\mu\nu} = \eta_{\mu\nu} + h_{\mu\nu} \quad ,
\qquad |h_{\mu\nu}| \ll 1 \ . \eeq

\noi To first order in $h_{\mu\nu}$ the Riemann tensor
$R_{\mu\nu\rho\sigma}$ reads

\beq \label{2.2e} R_{\mu\nu\rho\sigma} = {1 \over 2} \left (
h_{\mu\sigma , \nu \rho} + h_{\nu \rho , \mu \sigma} - h_{\mu \rho ,
\nu \sigma} - h_{\nu \sigma , \mu \rho} \right ) \ . \eeq

\noi $R_{\mu\nu\rho\sigma}$ has the symmetry properties

\beq \label{2.3e} R_{\mu\nu\rho\sigma} = - R_{\nu \mu \rho \sigma} = -
R_{\mu \nu \sigma \rho} = + R_{\rho \sigma \mu \nu} \ . \eeq

\noi It satisfies the first Bianchi identity (or cyclic identity)

\beq \label{2.4e} R_{\mu\nu\rho\sigma} + R_{\mu\sigma\nu\rho} +
R_{\mu\rho\sigma\nu} = 0 \eeq

\noi and the second Bianchi identity

\beq \label{2.5e} \partial_{\lambda} R_{\mu\nu\rho\sigma} +
\partial_{\rho} R_{\mu\nu \sigma\lambda} + \partial_{\sigma}
R_{\mu\nu\lambda\rho} = 0 \ . \eeq

\noi In addition we will impose, to start with, the vacuum equations of
motion which imply the vanishing of the Ricci tensor:

\beq \label{2.6e} R_{\phantom{\nu}\rho}^{\nu} \equiv
R_{\phantom{\mu\nu}\rho\mu}^{\mu\nu} = 0 \ . \eeq

\noi Now we define the dual Riemann tensor
$\widetilde{R}_{\mu\nu\rho\sigma}$ by

\beq \label{2.7e} \widetilde{R}_{\mu\nu\rho\sigma} = {1 \over 4} \left
[ \varepsilon_{\mu\nu\alpha\beta} \
R^{\alpha\beta}_{\phantom{\alpha\beta}\rho\sigma} +
R_{\mu\nu}^{\phantom{\mu\nu}\alpha\beta}\ 
\varepsilon_{\alpha\beta\rho\sigma} \right ] \ . \eeq

\noi In order to treat metrics with Euclidean and Lorentzian signatures
simultaneously in the following, we introduce the sign $\sigma$ with

\bea \label{2.8e} &&\sigma = + 1 \quad \hbox{(Euclidean signature)} \nn
\\ && \sigma = - 1 \quad \hbox{(Lorentzian signature)} \ . \eea

\noi Then the $\varepsilon$ tensor satisfies

\beq \label{2.9e} \varepsilon_{\mu\nu\rho\sigma} \
\varepsilon^{\mu\nu\rho\sigma} = 24 \sigma \ . \eeq

\noi The duality transformation (\ref{2.7e}) ensures that the dual
Riemann tensor is symmetric,

\beq \label{2.10e} \widetilde{R}_{\mu\nu\rho\sigma} =
\widetilde{R}_{\rho\sigma\mu\nu} \ . \eeq

\noi If we apply the duality transformation (\ref{2.7e}) twice, we
would like to obtain the identity (up to a sign depending on the
signature of the metric):

\beq \label{2.11e} \widetilde{\widetilde{R}}_{\mu\nu\rho\sigma}
\mathrel{\mathop =^{!}} \sigma \ R_{\mu \nu\rho\sigma} \ . \eeq

\noi One finds, however, that (\ref{2.11e}) holds only if
$R_{\mu\nu\rho\sigma}$ satisfies

\beq \label{2.12e} R_{\mu \nu\rho\sigma} = {\sigma \over 4} \
\varepsilon_{\mu\nu\alpha\beta} \ R^{\alpha\beta\gamma\delta} \
\varepsilon_{\gamma\delta\rho\sigma} \ . \eeq

\noi This equation is not an identity; if one replaces $R$ by
$\widetilde{R}$ on its left-hand side, it corresponds to what has been
denoted duality transformation in \cite{7r}. \par

Instead of (\ref{2.7e}) we could have defined dual Riemann tensors by

\beq \label{2.13e} \widetilde{R}_{\mu\nu\rho\sigma}^{(1)} = {1 \over 2}
\ \varepsilon_{\mu \nu \alpha\beta} \
R^{\alpha\beta}_{\phantom{\alpha\beta}\rho\sigma} \eeq

\noi or

\beq \label{2.14e} \widetilde{R}_{\mu\nu\rho\sigma}^{(2)} = {1 \over 2}
\ R_{\mu\nu}^{\phantom{\mu\nu}\alpha\beta}\ \varepsilon_{\alpha \beta
\rho\sigma}  \ . \eeq

\noi In both cases (\ref{2.11e}) is always satisfied; however, the
symmetry property (\ref{2.10e}) holds for $\widetilde{R}^{(1)}$ or
$\widetilde{R}^{(2)}$ only if $R$ satisfies (\ref{2.12e}). Thus, if we
require that both eqs. (\ref{2.10e}) and (\ref{2.11e}) hold, we have to
restrict ourselves to metrics whose Riemann tensor satisfies
(\ref{2.12e}), and now all duality transformations (\ref{2.7e}),
(\ref{2.13e}) and (\ref{2.14e}) are equivalent. For definiteness,
however, we will subsequently refer to eq. (\ref{2.7e}) as our
definition of the duality transformation. \par

The properties of $\widetilde{R}_{\mu\nu\rho\sigma}$ have previously
been discussed by Hull \cite{9r}. Its first Bianchi identity follows
from the vanishing of the Ricci tensor (\ref{2.6e}):

\beq \label{2.15e} \widetilde{R}_{\mu\nu\rho\sigma} +
\widetilde{R}_{\mu\sigma\nu\rho} + \widetilde{R}_{\mu\rho\sigma\nu} = 0
\ . \eeq

\noi Its second Bianchi identity follows from the second Bianchi
identity of $R_{\mu\nu\rho\sigma}$, (\ref{2.5e}), if (\ref{2.6e})
holds:

\beq \label{2.16e} \partial_{\lambda} \widetilde{R}_{\mu\nu\rho\sigma}
+ \partial_{\rho}\widetilde{R}_{\mu\nu\sigma\lambda} +
\partial_{\sigma} \widetilde{R}_{\mu\nu\lambda\rho} = 0 \ . \eeq

\noi Finally the first Bianchi identity for $R_{\mu\nu\rho\sigma}$,
eq. (\ref{2.4e}), implies the vanishing of the dual Ricci tensor:

\beq \label{2.17e} \widetilde{R}^{\nu}_{\ \rho} \equiv
\widetilde{R}^{\mu\nu}_{\phantom{\mu\nu}\rho\mu}  = 0 \ . \eeq

\noi $\widetilde{R}_{\mu\nu\rho\sigma}$ has thus the same properties as
$R_{\mu\nu\rho\sigma}$. Its symmetries together with the Bianchi
identities (\ref{2.15e}) and (\ref{2.16e}) are sufficient to prove that
$\widetilde{R}_{\mu\nu\rho\sigma}$ can be written in terms of a dual
linearized metric $\widetilde{h}_{\mu\nu}$ \cite{13r} (the vanishing of
the dual Ricci tensor is not needed to this end) as

\beq \label{2.18e} \widetilde{R}_{\mu\nu\rho\sigma} = {1 \over 2} \left
( \widetilde{h}_{\mu\sigma ,\nu\rho} + \widetilde{h}_{\nu\rho,
\mu\sigma} - \widetilde{h}_{\mu\rho , \nu\sigma} -
\widetilde{h}_{\nu\sigma, \mu \rho} \right ) \ . \eeq

\noi An explicit formula for $\widetilde{h}_{\mu\nu}$ in terms of
$\widetilde{R}_{\mu\nu\rho\sigma}$ is given in \cite{13r} in the
coordinate gauge $x^{\mu} \widetilde{h}_{\mu\nu} = x^{\nu}
\widetilde{h}_{\mu\nu} = 0$, which reads in our convention (\ref{2.2e})

\beq \label{2.19e} \widetilde{h}_{\mu\nu}(x) = - \int_0^1 dt \int_0^t
dt't' \ x^{\rho} x^{\sigma} \ \widetilde{R}_{\mu\rho\nu\sigma}(t'x) \ .
\eeq

\noi Thus the $S$-dual of linearized gravity can be constructed
explicitly. Let us close this short section with a note on gauge
symmetries. $R_{\mu\nu\rho\sigma}$ is invariant under

\beq \label{2.20e} h'_{\mu\nu} = h_{\mu\nu} + \partial_{\mu}
\Lambda_{\nu} + \partial_{\nu} \Lambda_{\mu} \ , \eeq

\noi and $\widetilde{R}_{\mu\nu\rho\sigma}$ is invariant under

\beq \label{2.21e} \widetilde{h}'_{\mu\nu} = \widetilde{h}_{\mu\nu} +
\partial_{\mu} \widetilde{\Lambda}_{\nu} + \partial_{\nu}
\widetilde{\Lambda}_{\mu} \ . \eeq

\noi At this level the gauge parameters $\Lambda_{\mu}$ and
$\widetilde{\Lambda}_{\mu}$ are independent (as the ``electric'' and
``magnetic'' U(1) gauge symmetries in the context of electromagnetic
duality) and, moreover, not seemingly related to general coordinate
transformations. Clearly this is an artefact of linearized gravity. In
full gravity one requires that both $R_{\mu\nu\rho\sigma}$ and
$\widetilde{R}_{\mu\nu\rho\sigma}$ transform as tensors under general
coordinate transformations, and the transformations of $h_{\mu\nu}$ and
$\widetilde{h}_{\mu\nu}$ (or better $g_{\mu\nu}$ and
$\widetilde{g}_{\mu\nu}$) get linked through the link (\ref{2.7e})
between the Riemann tensors (which remains, however, to be
covariantized).

\mysection{The Dual of Full Gravity and Taub-NUT Spaces}
\hspace*{\parindent} An attempt to covariantize the duality
transformation (\ref{2.7e}) and to preserve the properties of the dual
Riemann tensor $\widetilde{R}_{\mu\nu\rho\sigma}$ meets the following
obstacles:\par

a) in order to render the $\varepsilon$ tensor in (\ref{2.7e})
covariant we have to multiply it by $\sqrt{g}$ (or
$\sqrt{\widetilde{g}}$ under the assumption that $\widetilde{g}$ exists
and transforms as $g$); in order to raise/lower indices we have to
contract them with $g^{\mu\nu}$ (or $\widetilde{g}^{\mu\nu}$), notably
in order to derive the first Bianchi identity (\ref{2.15e}) for
$\widetilde{R}_{\mu\nu\rho\sigma}$ from (\ref{2.6e}). Either choice
between $g^{\mu\nu}$ or $\widetilde{g}^{\mu\nu}$ destroys the symmetry
corresponding to $g \leftrightarrow \widetilde{g}$ together with $R
\leftrightarrow \widetilde{R}$. \par

b) the derivatives in the second Bianchi identities (\ref{2.5e}) and
(\ref{2.16e}) have to be covariantized with the help of Christoffel
symbols which are {\it both} related to the {\it same} metric
$g_{\mu\nu}$ (or $\widetilde{g}_{\mu\nu}$), if (\ref{2.16e}) and
(\ref{2.5e}) are required to follow from each other. Hence it is not
possible to maintain the standard form of both covariantized Bianchi
identities (\ref{2.5e}) and (\ref{2.16e}), {\it and} to derive
(\ref{2.16e}) from (\ref{2.5e}) after covariantization. \par

The most convenient way to circumvent the obstacles a) is to define the
duality transformation (\ref{2.7e}) in terms of differential forms
\cite{10r}: One expresses the metric $g_{\mu\nu}$ in terms of an
orthonormal base of vierbeins $e_{\ \mu}^a$ (with inverses $e_a^{\
\mu}$),

\beq \label{3.1e} g_{\mu\nu} = \eta_{ab} \ e_{\ \mu}^a \ e_{\ \nu}^b
\quad , \qquad \eta_{ab} = g_{\mu\nu} \ e_a^{\ \mu} \ e_b^{\ \nu} \ ,
\eeq

\noi and defines the Riemann tensor with (latin) indices in the flat
tangent space:

\beq \label{3.2e} R_{abcd} = e^{\ \mu}_a \ e^{\ \nu}_b \ e_c^{\ \rho} \
e_d^{\ \sigma} \ R_{\mu\nu\rho\sigma} \ . \eeq

\noi Instead of eq. (\ref{2.7e}) we define now the dual Riemann tensor
by

\beq \label{3.3e} \widetilde{R}_{abcd} = {1 \over 4} \left [
\varepsilon_{abef} \ R_{\phantom{ef}cd}^{ef} + 
R_{ab}^{\phantom{ab}ef}\ \varepsilon_{efcd} \right ] \ . \eeq

\noi Since now indices are raised and lowered with the flat metric
$\eta_{ab}$ it is straightforward to show that the first Bianchi
identity for $\widetilde{R}_{abcd}$, the analog of eq. (\ref{2.15e}),
still follows from the vanishing Ricci tensor (and vice versa).  \par

The obstacle b), however, has not been removed: The covariant second
Bianchi identity for $R_{abcd}$ involves the spin connection associated
to the metric $g_{\mu\nu}$ \cite{10r}, and it is not possible to prove
a covariant second Bianchi identity for $\widetilde{R}_{abcd}$
involving a spin connection associated to a dual metric
$\widetilde{g}_{\mu\nu}$ (which is not even defined at this stage).
\par

Hence, although a dual Riemann tensor $\widetilde{R}_{abcd}$ can always
be constructed from eq. (\ref{3.3e}), it is not clear whether it can
always be derived from a dual metric $\widetilde{g}_{\mu\nu}$.  (An
iterative procedure which allows to extend the result (\ref{2.19e}) of
linearized gravity to higher orders in $h_{\mu\nu}$ (or
$\widetilde{h}_{\mu\nu}$) may possibly be derived along  the lines in
ref. \cite{14r}.) \par

Therefore it is of interest to find specific examples of metrics
$g_{\mu\nu}$ and $\widetilde{g}_{\mu\nu}$, whose Riemann tensors are
related through (\ref{3.3e}) without having to resort to a weak field
expansion. Subsequently we will consider Taub-NUT spaces for general
$m$ and $\ell$, which contain the Schwarzschild metric in the limit
$\ell \to 0$. For $m = \ell$, in Euclidean space-time, they are
well-known to be self-dual in the sense of eq. (\ref{3.3e})
\citd{11r}{10r}. \par

The Taub-NUT metric can be written as follows \citd{1r}{2r} (we recall
the definition of the sign $\sigma$ in eq. (\ref{2.8e})):

\beq \label{3.4e} ds^2 = \sigma f^2(r) \left ( dt + 4 \ell \sin^2
{\theta \over 2} d\phi \right )^2 + f^{-2}(r) dr^2 + \left ( r^2 -
\sigma \ell^2 \right ) \left ( d\theta^2 + \sin^2 \theta d\phi^2 \right
) \eeq

\noi with

\beq \label{3.5e} f^2(r) = 1 - {2(mr - \sigma \ell^2) \over r^2 -
\sigma \ell^2} \ . \eeq

\noi The non-vanishing components of the Riemann tensor $R_{abcd}$, as
defined in eq. (\ref{3.2e}), have been computed by Misner \cite{2r}:
\bea \label{3.6e} &&R_{0101} = - 2 A_{m,\ell}(r) \nn \\ &&R_{0202} =
R_{0303} = A_{m,\ell}(r) \nn \\ &&R_{1212} = R_{1313} = \sigma
A_{m,\ell}(r) \nn \\ &&R_{2323} = -2 \sigma A_{m,\ell}(r) \nn \\
&&R_{0312} = - R_{0213} =  D_{m,\ell}(r) \nn \\ &&R_{0123} =
- 2 D_{m,\ell}(r) \eea

\noi with
\bea \label{3.7e} &&A_{m, \ell }(r) ={-\sigma m r^3 + 3 \ell^2 r^2 - 3m
\ell^2 r + \sigma \ell^4 \over (r^2 - \sigma \ell^2)^3} \ ,  \nn \\
&&D_{m, \ell }(r) ={\sigma \ell r^3 - \sigma \ell mr^2 + 3 r \ell^3 -
m\ell^3  \over (r^2 - \sigma \ell^2)^3} \ . \eea

\noi One easily verifies that $R_{abcd}$ satisfies the analog of eq.
(\ref{2.12e}) (with greek indices replaced by latin indices), hence
eqs. (\ref{2.10e}) and (\ref{2.11e}) are both satisfied, and we can
replace eq. (\ref{3.3e}) by the analogs of eqs. (\ref{2.13e}) or
(\ref{2.14e}). Subsequently we have to specify one component of the
tensor $\varepsilon_{abcd}$: we use, both for Euclidean and Lorentzian
signatures,

\beq \label{3.8e} \varepsilon_{0123} = + 1 \ . \eeq

\noi Then we obtain the following non-vanishing components of
$\widetilde{R}_{abcd}$:

\bea \label{3.9e} &&\widetilde{R}_{0101} = - 2 D_{m,\ell}(r) \nn \\
&&\widetilde{R}_{0202} = D_{m,\ell}(r) \nn \\ &&\widetilde{R}_{1212} =
\widetilde{R}_{1313} = \sigma D_{m,\ell}(r) \nn \\
&&\widetilde{R}_{2323} = -2 \sigma D_{m,\ell}(r) \nn \\
&&\widetilde{R}_{0312} = - \widetilde{R}_{0213} =  \sigma A_{m,\ell}(r)
\nn \\ &&\widetilde{R}_{0123} =  - 2 \sigma A_{m,\ell}(r) \ . \eea

\noi The search for a dual metric, which is associated to the dual
Riemann tensor $\widetilde{R}_{abcd}$, is greatly simplified by the
following properties of the functions $A_{m,\ell}(r)$ and
$D_{m,\ell}(r)$: If one defines

\beq \label{3.10e} m' = \sigma \ {\ell^2 \over m} \eeq

\noi one has
\bea \label{3.11e} &&A_{m',\ell}(r) = - \sigma \ {\ell \over m} \ D_{m,
\ell}(r) \ , \nn \\ &&D_{m',\ell}(r) = {\ell \over m} \ A_{m, \ell}(r)
\ . \eea

\noi Furthermore a rescaling of the Riemann tensor corresponds to a
rescaling of the metric:

\beq \label{3.12e} \gamma R_{abcd} = R_{abcd} \Big
|_{g_{\mu\nu} \to \gamma^{-1} g_{\mu\nu}} \ . \eeq

\noi Inserting eqs. (\ref{3.11e}) and (\ref{3.12e}) into (\ref{3.9e})
one finds that the following metric generates a Riemann tensor whose
components are given by (\ref{3.9e}):

\beq \label{3.13e} \widetilde{ds}^2 = {\ell \over m} \left \{ -
\widehat{f}^{\, 2}(r) \left ( dt + 4 \ell \sin^2 {\theta \over 2} d\phi
\right )^2 - \sigma \left [ \widehat{f}^{-2} (r) dr^2 + \left ( r^2 -
\sigma \ell^2 \right ) \left (d \theta^2 + \sin^2 \theta d\phi^2 \right )
\right ] \right \} \eeq

\noi with

\beq \label{3.14e} \widehat{f}^{\, 2} (r) = 1 + {2 \sigma \ell^2 (m - r)
\over m (r^2 - \sigma \ell^2)} \ . \eeq

\noi In order to bring the metric (\ref{3.13e}) into the form of the
metric (\ref{3.4e}) one has to rescale the coordinates $t$ and $r$,

\beq \label{3.15e} t = \sqrt{{m \over \ell}} \ t' \qquad , \qquad r =
\sqrt{{m \over \ell}} \ r' \ , \eeq

\noi and define

\beq \label{3.16e} \widetilde{m} = \sigma \ell^{5/2} m^{-3/2} \qquad
, \quad \widetilde{\ell} = \ell^{3/2} m^{-1/2} \ . \eeq

\noi Now $\widetilde{ds}^2$ becomes

\beq \label{3.17e} \widetilde{ds}^2 = - \widetilde{f}^2(r') \left ( dt'
+ 4 \widetilde{\ell} \sin^2 {\theta \over 2} d \phi \right )^2 - \sigma
\left [ \widetilde{f}^{-2}(r') dr^{'2} + \left ( r^{'2} - \sigma
\widetilde{\ell}^2 \right ) \left ( d \theta^2 + \sin^2 \theta d\phi^2
\right ) \right ] \eeq

\noi with

\beq \label{3.18e} \widetilde{f}^2(r') = 1 - {2(\widetilde{m} r' -
\sigma \widetilde{\ell}^2) \over r^{'2} - \sigma \widetilde{\ell}^2} \
. \eeq

\noi Now $\widetilde{f}^2$ is of the same for as $f^2$ in eq.
(\ref{3.5e}), but $\widetilde{ds}^2$ still differs by a factor $-
\sigma$ from $ds^2$ in eq. (\ref{3.4e}). In the Euclidean case $(\sigma =
+1)$ we find that, for $m = \ell$, the metric and hence the Riemann
tensor are anti-self-dual. Here we have tacitely assumed that $m$ and
$\ell$ are positive. For $\ell$ negative and $m = - \ell$, however, we
have to replace $\ell$ by $- \ell$ in eqs. (\ref{3.15e}) and
(\ref{3.16e}), and one finds that the metric and the Riemann tensor are
self-dual. In the Lorentzian case ($\sigma = - 1$) eqs. (\ref{3.16e})
imply that $\widetilde{m}$ is negative if $\ell$ and $m$ are real and
positive. These are the well-known (anti-) self-duality properties of
Taub-NUT metrics with $|m| = |\ell|$. \par

Actually, up to rescalings of the $t$ and $r$ coordinates, only the
ratio $|m/\ell|$ has a physical significance in Taub-NUT spaces. From
eqs. (\ref{3.16e}) one finds immediately

\beq \label{3.19e} \left | {m \over \ell} \right | = \left |
{\widetilde{\ell} \over \widetilde{m}} \right | \ , \eeq

\noi hence $\ell$ and $m$ indeed exchange their role under the duality
transformation (\ref{3.3e}). This agrees with the interpretation of
$\ell$ as a ``magnetic'' mass \citm{3r}{7r}, and with the
interpretation of (\ref{3.3e}) as the analog of $S$-duality in Maxwell
theory. \par

However, the presence of the factors $\ell$ in eqs. (\ref{3.13e}) and
(\ref{3.15e}) shows that the analogy breaks down in the limit $\ell \to
0$: Now $\widetilde{ds}^2$ vanishes, or the rescaling (\ref{3.15e})
becomes singular. Hence the duals of the Schwarzschild metric (and,
similarly, the dual of the Taub-NUT metric with $m = 0$) do not exist.
\par

One could be tempted to confine oneself to the asymptotic regime $r \to
\infty$, and to apply the explicit formula (\ref{2.19e}) for
$\widetilde{h}_{\mu\nu}$ in the pure Schwarzschild case. Here one has
to use Cartesian coordinates instead of spherical coordinates such
that $|h_{\mu\nu}| \ll 1$ for $r \to \infty$. Then one finds indeed
that

\beq \label{3.20e} x^{\rho} x^{\sigma} \widetilde{R}_{\mu \rho
\nu\sigma} (t'x) = 0 \eeq

\noi for $\ell = 0$ in eq. (\ref{2.19e}), in agreement with the results
above. \par

Hence the analogy between gravitational and Maxwell $S$-duality holds
only for ``dyonic'' Taub-NUT metrics with both $m$ and $\ell \not= 0$.

\mysection{Duality in Taub-NUT-de Sitter Spaces} \hspace*{\parindent}
In Taub-NUT-de Sitter spaces the Ricci tensor vanishes no longer, but
is proportional to a cosmological constant: 

\beq \label{4.1e} R_{\ c}^b \equiv R_{\phantom{ab}ca}^{ab} = \Lambda 
\delta_{\ c}^b \eeq

\noi instead of eq. (\ref{2.6e}). (We continue to work with tensors
with indices in flat tangent space, which are related to the tensors
with indices in ordinary space-time through the vierbeins $e_{\mu}^a$.
Furthermore we find it more convenient to define the cosmological
constant in terms of the Ricci tensor instead of the more conventional
definition in terms of the Einstein tensor.) 

Of course the first and second Bianchi identities for
$R_{abcd}$ remain intact, but now one has to wonder how one can obtain
the validity of the first Bianchi identity (\ref{2.15e}) for a dual
Riemann tensor which previously required the vanishing of the Ricci
tensor. \par

In some sense the cosmological constant $\Lambda$ represents a
(trivial) ``matter'' degree of freedom, and quite generally duality in
the presence of matter (if it exists at all) requires some mixing
between the ``gauge fields'' and ``matter''.
\par

First, the dual of a cosmological constant (in a sense specified below)
turns out to be a three-form field $A_{abc} = A_{[abc]}$, with a field
strength

  \beq \label{4.2e}
  F_{abcd} = \partial_{[a} A_{bcd]} \ .
\eeq

\noi The equations of motion for $A_{abc}$ read

  \beq \label{4.3e}
  \partial^a F_{abcd} = 0 \ ,
\eeq

\noi and the only solutions respecting Lorentz covariance are of the
 form

  \beq \label{4.4e}
  F_{abcd} = \Sigma \varepsilon_{abcd} \qquad , \qquad \Sigma = 
\hbox{const.} \ .
\eeq

\noi Now we consider the following generalization of the duality 
transformation (\ref{3.3e}):

\bminiG{4.5e}
\label{4.5ae}
\widetilde{R}_{abcd} = {1 \over 4} \left [ \varepsilon_{abef} \left ( 
R_{\phantom{ef}cd}^{ef} + F^{ef}_{\phantom{ef}cd} \right ) + \left ( 
R_{ab}^{\phantom{ab}ef} + F_{ab}^{\phantom{ab}ef} \right )\
\varepsilon_{efcd} \right ] + {1 \over 12} \  \varepsilon_{abcd} R \ ,
  \eeeq
\beeq
  \label{4.5be}
  \widetilde{F}_{abcd} = - {1 \over 12} \ \varepsilon_{abcd} R \ ,
  \emini

\noi where

  \beq \label{4.6e}
  R \equiv R^{ab}_{\phantom{ab}ba} \ .
\eeq

\noi Let us discuss the properties of $\widetilde{R}_{abcd}$. First,
$\widetilde{R}_{abcd}$ still has the same symmetry properties
(\ref{2.3e}) as $R_{abcd}$. Next, the first Bianchi identity still
holds:

\beq \label{4.7e} \widetilde{R}_{abcd} + \widetilde{R}_{adbc} +
\widetilde{R}_{acdb} = 0 \eeq

\noi where one has to use eq. (\ref{4.4e}) for $F_{abcd}$ (i.e. the
equation of motion for $A_{abc}$), and the last term $\sim R$ in
(\ref{4.5ae}) serves to cancel the contributions proportional to the
cosmological constant. Also, the second Bianchi identity still holds at
the linearized level:

\beq \label{4.8e} \partial_e \widetilde{R}_{abcd} + \partial_c
\widetilde{R}_{abde} + \partial_d \widetilde{R}_{abec} = 0 \eeq

\noi where one has to use the linearized second Bianchi identity for
$R_{abcd}$, and the fact that both the Ricci tensor $R_{\ b}^a$ and
$F_{abcd}$ are constant. Eqs. (\ref{4.7e}) and (\ref{4.8e}) are already
sufficient to prove that, at the linearized level,
$\widetilde{R}_{abcd}$ can again be expressed in terms of a dual metric
$\widetilde{h}_{ab}$ as in eq. (\ref{2.19e}) (the distinction between
latin and greek indices is meaningless at the linearized level). \par

For the dual Ricci tensor one obtains

\beq \label{4.9e} \widetilde{R}_{\ b}^a = - 3 \sigma \Sigma \
\delta_{\ b}^a
\eeq

\noi with the help of the first Bianchi identity for $R_{abcd}$, and
eq. (\ref{4.4e}) for $F_{abcd}$. Hence $\widetilde{R}_{\ b}^a$ is
proportional to a dual cosmological constant $\widetilde{\Lambda}$ with

\beq \label{4.10e} \widetilde{\Lambda} = - 3 \sigma \Sigma \ . \eeq

$\widetilde{F}_{abcd}$ always satisfies the Bianchi identity

\beq \label{4.11e}
\partial_{[a} \widetilde{F}_{bcde]} = 0
\eeq

\noi which is a trivial identity in $d = 4$. The dual equations of motion

\beq \label{4.12e}
\partial^a \widetilde{F}_{abcd} = 0
\eeq

\noi follow from the constancy of the Riemann scalar $R$: together
with (\ref{4.1e}) eq. (\ref{4.5be}) gives evidently

\beq \label{4.13e}
\widetilde{F}_{abcd} = - {1 \over 3} \varepsilon_{abcd}  \Lambda \ .
\eeq

\noi Eq. (\ref{4.11e}) shows that $\widetilde{F}_{abcd}$ can be
written as

\beq \label{4.14e}
\widetilde{F}_{abcd} = \partial_{[a} \widetilde{A}_{bcd]}
\eeq

\noi and the solution of the equation of motion (\ref{4.12e}) for 
$\widetilde{A}_{abc}$ gives

\beq \label{4.15e}
\widetilde{F}_{abcd} = \varepsilon_{abcd} \widetilde{\Sigma}
\eeq

\noi with, from (\ref{4.13e}),

\beq \label{4.16e}
\widetilde{\Sigma} = - {1 \over 3} \Lambda \ .
\eeq

\noi Equations (\ref{4.10e}) and (\ref{4.16e}) clarify in what
sense $A_{abc}$ is dual to the cosmological constant: Up to a factor
$(-3)$ (and the sign $\sigma$) the duality transformations (\ref{4.5e})
lead to an interchange of $\Sigma$, the parameter characterizing the
solution of the equation of motion of $A_{abc}$, with the cosmological
constant $\Lambda$. \par

The effect of a double duality transformation on $F_{abcd}$ is easily
obtained from eqs. (\ref{4.13e}) and (\ref{4.10e}):

\beq \label{4.17e} \widetilde{\widetilde{F}}_{abcd} = \sigma F_{abcd}
\ . \eeq

\noi After some calculation one finds that the effect of a double
duality transformation on $R_{abcd}$ is the same as before:

\beq \label{4.18e} \widetilde{\widetilde{R}}_{abcd} = \sigma R_{abcd}
\eeq

\noi if $R_{abcd}$ satisfies

\beq \label{4.19e} R_{abcd} = {\sigma \over 4} \varepsilon_{abef} \
R^{efgh} \varepsilon_{ghcd} \ . \eeq

\noi Hence, on metrics which satisfy (\ref{4.19e}), our generalized
duality transformation (\ref{4.5e}) has all desirable properties. As
before, however, the validity of a second Bianchi identity for
$\widetilde{R}_{abcd}$ can not be proven beyond the linearized level.
\par

Let us now study the effect of (\ref{4.5e}) on Taub-NUT-de Sitter
metrics. These metrics can be written in the same form as the Taub-NUT
metric (\ref{3.4e}); it suffices to replace the function $f^2(r)$ by

\beq \label{4.20e} f^2(r) = 1 -{2\left ( mr - \sigma \ell^2 \right ) -
\Lambda \left ( {1 \over 3} r^4 - 2 \sigma \ell^2 r^2 - \ell^4 \right )
\over r^2 - \sigma \ell^2} \ . \eeq

\noi Now the non-vanishing components of the Riemann tensor are,
instead of eqs. (\ref{3.6e}),
\bea \label{4.21e} &&R_{0101} = - 2 A_{\Lambda} (r) \nn \\
&&R_{0202} = R_{0303} = C_{\Lambda}(r) \nn \\
&&R_{1212} = R_{1313} = \sigma C_{\Lambda} (r) \nn \\ 
&&R_{2323} = - 2 \sigma 
A_{\Lambda} (r) \nn \\ &&R_{0312} = - R_{0213} = D_{\Lambda} (r) \nn \\
&&R_{0123} = - 2 D_{\Lambda} (r) \eea

\noi with
\bea \label{4.22e} &&A_{\Lambda} (r) = \left ( 1 - {4 \over 3}  \sigma
\Lambda \ell^2 \right ) A_{\bar{m}, \ell} (r) + {\sigma \over 6} 
\Lambda \nn \\ &&C_{\Lambda} (r) = \left ( 1 - {4 \over 3} \sigma
\Lambda \ell^2 \right ) A_{\bar{m}, \ell} (r) - {\sigma \over 3} 
\Lambda \nn \\ &&D_{\Lambda} (r) = \left ( 1 - {4 \over 3} \sigma
\Lambda \ell^2 \right ) D_{\bar{m}, \ell} (r) \eea

\noi where $A_{\bar{m}, \ell}$ and $D_{\bar{m}, \ell}$ are the
functions defined in eq. (\ref{3.7e}) and

\beq \label{4.23e} \bar{m} = m \left ( 1 - {4 \over 3} \sigma \Lambda
\ell^2 \right )^{-1} \ . \eeq

\noi Constructing the components of the dual Riemann tensor from eq.
(\ref{4.5ae}) one obtains now additional contributions from the terms
$\sim F_{abcd}$ and $\sim R$. One finds
\bea \label{4.24e}
&&\widetilde{R}_{0101} = - 2D_{\Lambda} + \Sigma \nn \\
&&\widetilde{R}_{0202} = \widetilde{R}_{0303} = D_{\Lambda} + \Sigma \nn \\
&&\widetilde{R}_{1212} = \widetilde{R}_{1313} = \sigma D_{\Lambda} + 
\sigma  \Sigma \nn \\
&&\widetilde{R}_{2322} = - 2 \sigma D_{\Lambda} + \sigma \Sigma \nn \\
&&\widetilde{R}_{0312} = \widetilde{R}_{0213} = \sigma C_{\Lambda} + 
{1 \over 3} \Lambda \nn \\
&&\widetilde{R}_{0123} = -2 \sigma A_{\Lambda} +  {1 \over 3} \Lambda \ .
\eea

\noi Again the properties (\ref{3.11e}) of the functions $A(r)$, $D(r)$
help to find a metric which reproduces the same components of the
Riemann tensor:

\bea
\label{4.25e}
&&\widetilde{ds}^2 = - {\ell \over m + 4 \sigma \Sigma \ell^3} \left 
\{ \widehat{f}^{\, 2}(r) \left ( dt
+ 4 \ell \sin^2 {\theta \over 2} d \phi \right )^2 + \sigma
\left [ \widehat{f}^{-2}(r) dr^{2} \right . \right . \nn \\
&&\left . \left . + \left ( r^{2} - \sigma
\ell^2 \right ) \left ( d \theta^2 + \sin^2 \theta d\phi^2
\right ) \right ] \right \}
\eea

\noi with

\beq \label{4.26e} \widehat{f}^{\, 2}(r) = 1 + {2 \sigma \ell^2 \left ( m + 4
\sigma \Sigma \ell^3 - r\left ( 1 - {4 \over 3} \sigma \Lambda \ell^2
\right ) \right ) + 3 \Sigma \ell \left ( {1 \over 3}r^4 - 2 \sigma r^2
\ell^2 - \ell^4 \right ) \over \left ( m + 4 \sigma \Sigma \ell^3
\right ) \left ( r^2 - \sigma \ell^2 \right )} \ . \eeq

\noi A rescaling similar to eqs. (\ref{3.15e}),

\beq \label{4.27e} t = \sqrt{{m + 4 \sigma \Sigma \ell^3 \over \ell}} \
t' \qquad , \qquad r = \sqrt{{m + 4 \sigma \Sigma \ell^3 \over \ell}} \
r' \ , \eeq

\noi and dual parameters similar to eqs. (\ref{3.16e}),

\bea \label{4.28e} &&\widetilde{m} = \sigma \left ( 1 - {4 \over 3} 
\sigma\Lambda \ell^2 \right )\left ( m + 4 \sigma \Sigma
\ell^3 \right )^{-3/2} \ell^{5/2} \quad , \quad \widetilde{\ell} = \ell^{3/2}
\left ( m + 4 \sigma \Sigma \ell^2 \right )^{-{1 \over 2}} \ ,\nn \\
&&\widetilde{\Lambda} = - 3 \sigma \Sigma \ , \eea

\noi allow to write the dual metric again in the form (\ref{3.17e})
with

\beq \label{4.29e} \widetilde{f}^2 (r') = 1 - {2\left ( \widetilde{m}
r' - \sigma \widetilde{\ell}^2 \right ) + \sigma \widetilde{\Lambda}
\left ( {1 \over 3}r^{'4} - 2 \sigma r^{'2} \widetilde{\ell}^2 -
\widetilde{\ell}^4 \right ) \over r^{'2} - \sigma \widetilde{\ell}^2} \
. \eeq

\noi Thus, up to signs (cf. the remarks below eq. (\ref{3.18e})) the
metric dual to a Taub-NUT-de Sitter metric is again of the Taub-NUT-de
Sitter form. Now the limit $m \to 0$ actually exists, but the limit
$\ell \to 0$ (pure Schwarzschild-de Sitter) is still singular. \par

Let us close this section with some remarks on a dual action. In the
present approach we have obtained simple duality relations between the
cosmological constant $\Lambda$ and $\Sigma$, cf. eqs. (\ref{4.10e}),
(\ref{4.16e}) and (\ref{4.28e}). Clearly the equation of motion for
$A_{abc}$, eq. (\ref{4.3e}), follows from a quadratic ``matter'' action
$\sim \lambda F_{abcd}F^{abcd}$. Then the cosmological constant
$\Lambda$ is actually equal to $24\lambda \sigma \Sigma^2$ plus an
arbitrary additional contribution $\Lambda '$. The same holds for the
dual cosmological constant $\widetilde{\Lambda}$. It is then
straightforward to write down duality relations between $\Lambda '$,
$\Sigma$ and $\widetilde{\Lambda}'$, $\widetilde{\Sigma}$ (which depend
on the couplings $\lambda$, $\widetilde{\lambda}$), but we did not find
these relations very illuminating. The previous relations show in a
much more direct way that, e.g., $\widetilde{\Lambda} = 0$ iff $\Sigma
= 0$ (and vice versa) which may have some interesting applications.

\mysection{Conclusions and Outlook} \hspace*{\parindent}
In the present paper we tried to push the concept of gravitational
$S$-duality beyond linearized gravity. We have seen that the duality
relation between the first Bianchi identity and the equations of motion
can be maintained. It is then of interest to find out, under which
conditions the dual Riemann tensor satisfies the second Bianchi identity
or, equivalently, under which conditions a dual metric exists. At least
to this end it is very helpful to study explicitly the properties of
metrics related by $S$-duality. \par

In the case of Taub-NUT metrics we have seen that, as expected,
$S$-duality corresponds to an exchange of the parameters $m$ and
$\ell$. However we have also seen that, due to the required rescaling
of the coordinates, the metric has to be ``dyonic'' in the sense that
both parameters $m$ and $\ell$ are non-vanishing. \par

Then we managed to generalize the concept of $S$-duality to Taub-NUT-de
Sitter spaces with cosmological constant $\Lambda$. Now a three-form
$A_{abc}$ has to be introduced. Somewhat unexpectedly it is not needed
to cancel the cosmological constant as proposed in \cite{12r}, but it
turns out to be {\it dual} to the cosmological constant. \par

This phenomenon may pave the way towards a new solution of the
cosmological constant problem: If, for some reason, this three-form
field vanishes such that $\Sigma = 0$ and, simultaneously, matter
couples to the {\it dual} metric $\widetilde{g}_{\mu\nu}$, the universe
as described by the dual metric has automatically a vanishing
cosmological constant $\widetilde{\Lambda} \sim \Sigma = 0$ (or vice
versa). \par

However, many properties of gravitational $S$-duality have to be better
understood before such a speculation can be supported more seriously.
The first open question is evidently for which metrics, which solve the
vacuum Einstein equations, a dual metric exists (beyond the linearized
level). The next open question concerns the coupling of gravity to
matter. We have already emphasized that, in the presence of matter, the
duality transformation rules certainly have to be modified. Our results
in the presence of a cosmological constant and/or a three-form field
indicate a first step in this direction. It may turn out, however, that
the framework of Riemannian geometry (with, e.g., the absence of
torsion) is too restrictive in order to allow for general gravitational
$S$-duality beyond a few particular configurations of metrics and
matter fields. \par

In the present approach to gravitational $S$-duality four space-time
dimensions evidently play a particular role: in $d \not= 4$ the dual
of the Riemann tensor has no longer the same number of indices. In
linearized gravity it can still be written in terms of a gauge field
with mixed symmetry properties \cite{9r}, but the relation of this
field to a Riemannian metric is not clear. If this relation could be
better understood, gravitational duality in the presence of matter in
$d = 4$ could be obtained from pure gravitational duality in $D > 4$
after compactification. \par

\vskip 20 truemm 
\noindent {\large\bf Acknowledgement} \par
\noindent It is a pleasure to thank B. Carter and M. Dubois-Violette for 
stimulating discussions.

\newpage \def\labelenumi{[\arabic{enumi}]} \noindent {\large\bf
References} \ben \item\label{1r} A. Taub, Ann. Math. {\bf 53} (1951)
472; \\ E. Newman, L. Tamburino, T. Unit, J. Math. Phys. {\bf 4}
(1963) 915.

\item\label{2r} C. Misner, J. Math. Phys. {\bf 4} (1963) 924.

\item\label{3r} J. Dowker, J. Roche, Proc. Phys. Soc. {\bf 92} (1967)
1; \\ R. Mignani, Lett. Nuovo Cim. {\bf 22} (1978) 597; \\
S. Ramaswamy, A. Sen, J. Math. Phys. {\bf 22} (1981) 2612; \\
G. Murphy, Int. J. Theor. Phys. {\bf 22} (1983) 477; \\
A. Zee, Phys. Rev. Lett. {\bf 55} (1985) 2379, Erratum ibid. {\bf 56} 
(1986) 1101; \\
D. Lynden-Bell, M. Nouri-Zonoz, Rev. Mod. Phys. {\bf 70} (1998) 427.

\item\label{4r} R. Geroch, J. Math. Phys. {\bf 12} (1971) 918.

\item\label{5r} A. Magnon, J. Math. Phys. {\bf 28} (1987) 2149; Gen.
Rel. Grav. {\bf 19} (1987) 809.

\item\label{6r} H. Garcia-Compean, O. Obregon, J. Plebanski, C.
Ramirez, Phys. Rev. {\bf D57} (1998) 7501; \\ H. Garcia-Campean, O.
Obregon, C. Ramirez, Phys. Rev. {\bf D58} (1998) 104012.

\item\label{7r} N. Dadhich, Mod. Phys. Lett. {\bf A14} (1999) 337;
ibid. 759; \\ M. Nouri-Zonoz, N. Dadhich, D. Lynden-Bell, Class.
Quant. Grav. {\bf 16} (1999) 1021.

\item\label{8r} J. Nieto, Phys. Lett. {\bf A262} (1999) 274.

\item\label{9r} C. Hull, Nucl. Phys. {\bf B583} (2000) 237; \\ C. Hull,
JHEP {\bf 0109} (2001) 27.

\item\label{10r} T. Eguchi, P. Gilkey, A. Hanson, Phys. Rept. {\bf 66}
(1980) 213.

\item\label{11r} S. Hawking, Phys. Lett. {\bf 60A} (1977) 81.

\item\label{12r} A. Aurilia, H. Nicolai, P. Townsend, Nucl. Phys. {\bf
B176} (1980) 509; \\ M. Duff, P. van Nieuwenhuizen, Phys. Lett. {\bf
B94} (1980) 179; \\ S. Hawking, Phys. Lett. {\bf B134} (1984) 403;\\
J. Brown, C. Teitelboim, Phys. Lett. {\bf B195} (1987) 177; \\ M.
Duff, Phys. Lett. {\bf B226} (1989) 36.

\item\label{13r} M. Dubois-Violette, M. Henneaux, Lett. Math. Phys.
{\bf 49} (1999) 245; \\ M. Dubois-Violette, M. Henneaux,
math-qa/0110088. 

\item\label{14r} S. Deser, Gen. Rel. Grav. {\bf 1} (1970) 9.

\een

\end{document}